\documentclass[aps,prb,superscriptaddress,reprint,showpacs,floatfix]{revtex4-1}

\usepackage[T1]{fontenc} 

\usepackage{amsmath}
\usepackage{amsthm}
\usepackage{amssymb}
\usepackage{amsfonts}
\usepackage{mathtools}
\usepackage{bm} 
\usepackage{mathrsfs}
\usepackage{physics}

\usepackage{soul} 
\usepackage{comment} 
\usepackage{chngcntr} 
\usepackage[colorinlistoftodos]{todonotes} 
\usepackage{natbib} 

\usepackage{chemformula} 

\usepackage{tabularx}
\usepackage{float}
\usepackage{graphicx}
\usepackage{booktabs}

\usepackage[colorlinks=true, citecolor=blue, linkcolor=blue, urlcolor=blue]{hyperref}

\usepackage{siunitx}
\AtBeginDocument{\RenewCommandCopy\qty\SI}
\DeclareSIUnit\angstrom{\text{Å}}

\begin{document}
\title{Wetting and Strain Engineering of 2D Materials on Nanopatterned Substrates}
\author{Davoud~Adinehloo}
\affiliation{Department of Electrical Engineering, University at Buffalo, Buffalo, NY 14228, USA}

\author{Joshua~R.~Hendrickson}
\affiliation{Sensors Directorate, Air Force Research Laboratory, Wright-Patterson AFB, Ohio 45433, USA}

\author{Vasili~Perebeinos}
\email{vasilipe@buffalo.edu}
\affiliation{Department of Electrical Engineering, University at Buffalo, Buffalo, NY 14228, USA}

\begin{abstract}

The fascinating realm of strain engineering and wetting transitions in two-dimensional (2D) materials takes place when placed on a two-dimensional array of nanopillars or one-dimensional rectangular grated substrates. Our investigation encompasses a diverse set of atomically thin 2D materials, including transition metal dichalcogenides, hexagonal boron nitride, and graphene, with a keen focus on the impact of van der Waals adhesion energies to the substrate on the wetting/dewetting behavior on nanopatterned substrates. We find a critical aspect ratio of the nanopillar or grating heights to the period of the pattern when the wetting/dewetting transition occurs. Furthermore, energy hysteresis analysis reveals dynamic detachment and re-engagement events during height adjustments, shedding light on energy barriers of 2D monolayer transferred on patterned substrates. Our findings offer avenues for strain engineering in 2D materials, leading to promising prospects for future technological applications.

\end{abstract}

\maketitle
\section{Introduction}
Two-dimensional (2D) materials have garnered immense attention due to their exceptional electronic~\cite{wang2018colloquium,wang2012electronics,avouris2007carbon}, optical~\cite{ma2021tunable,you2018nonlinear}, and mechanical~\cite{akinwande2017review,androulidakis2018tailoring} properties. These atomically thin materials exhibit remarkable promise for various applications, from high-performance electronics to advanced photonics~\cite{lemme20222d,ponnusamy2022emerging}. Particularly fascinating is their ability to function as single-photon emitters (SPEs), a role that holds great promise in the field of quantum information science and technology~\cite{aharonovich2016solid,koenderink2015nanophotonics,eisaman2011invited}.

In the realm of quantum technology, the quest for SPEs has been ongoing, and 2D materials, such as transition metal dichalcogenides (TMD) and hexagonal boron nitride (hBN), have emerged as promising candidates to fulfill this role~\cite{abramov2023photoluminescence,tonndorf2015single,he2015single,mendelson2021identifying}. Localization of excitons within 2D materials, induced by strain or defects, has been identified as the primary source of such photon emission~\cite{abramov2023photoluminescence,mendelson2021identifying,montblanch2021confinement,kim2022high,linhart2019localized}. Several articles have reported local strain-induced SPEs, providing valuable insight into their behavior~\cite{branny2017deterministic,liu2017ultra,palacios2017large}.

In fact, it is of paramount importance to highlight that the intricate microscopic mechanisms that govern the formation of strain-induced SPEs in 2D materials remain a topic of ongoing investigation~\cite{azzam2021prospects}. Further elucidation of these mechanisms is vital to advance our understanding of SPEs in 2D materials, with direct implications for the development of reliable and efficient SPEs integrated into quantum technologies~\cite{an2022perspectives,tian2016optoelectronic,carter2013quantum,palacios2017large}. 
The mechanical properties of 2D materials have been extensively explored~\cite{bertolazzi2011stretching}. These properties play an important role in shaping the electronic and optoelectronic performance of 2D materials~\cite{mahmud2023topological,lei2022graphene,touski2021vertical,10284770,hasani2023strain}. The adhesion energies of 2D materials to substrates, for example, have been reported in various materials: 6 to 28 meV/$\text{\AA}$ for graphene on silicon oxide~\cite{nguyen2011semiconducting,boddeti2013mechanics,boddeti2013mechanics,na2014ultra,koenig2011ultrastrong,akinwande2017review}, 9.4 meV/$\text{\AA}$ for graphene on silicon~\cite{dai2020mechanics,zong2010direct}, 1.12 meV/$\text{\AA}$ for MoS$_2$ on PMDS~\cite{brennan2015interface}, 6.3 meV/$\text{\AA}$ for MoS$_2$ on Al$_2$O$_3$~\cite{sanchez2018mechanics}, and 5 meV/$\text{\AA}$ for MoS$_2$ on SiO$_2$~\cite{sanchez2018mechanics}. Studies on the folding behavior of 2D materials~\cite{zhao2015two}, and investigations of heterostructures involving 2D materials~\cite{yu2021designing}, provide valuable information on their mechanical characteristics. Furthermore, specific structural arrangements have been examined, such as a graphene membrane stacked on triangular nanopillars~\cite{milovanovic2019strain}, and bubble structures~\cite{blundo2022vibrationa,blundo2021experimental,chen2022strain,lloyd2017adhesion,gastaldo2023tunable}. However, it is essential to note that while several studies have addressed the mechanical behavior of 2D materials, there is still a gap in our understanding regarding factors such as geometrical configuration, interfacial adhesion energy, and the period of the patterned structure, which warrants further investigation.

\begin{figure*}[t!]
\includegraphics[height=8   cm]{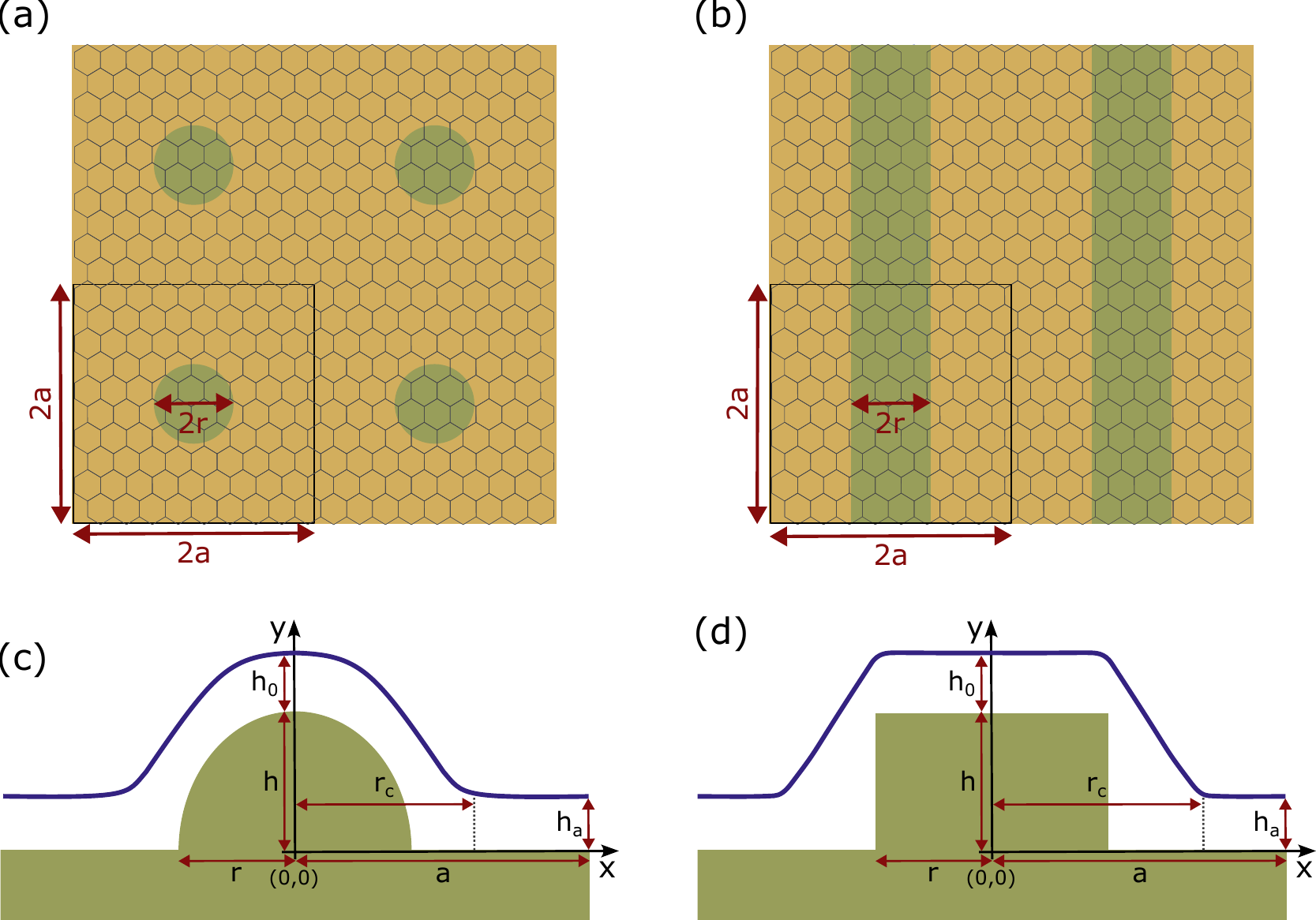}
\caption{\label{fig:Fig1} Schematic representations of two structural configurations: (a) a two-dimensional array of rounded nanopillars and (b) a rectangular one-dimensional substrate grating. Both nanopatterned substrates are coated with a monolayer of 2D material. (c) and (d) provide the vertical cross-sectional view of the 2D material when placed on nanopillars and rectangular grating, respectively. The key parameters include the period $2a$ of the patterned substrate, $h$ for the height of the nanostructure, $2r$ representing the diameter of the nanopillar or the width of the grating, and $h_a$ denoting the height of the 2D material at the edge of the unit cell at $x=a$.}
\end{figure*}

In this study, we offer a significant step towards advancing our ability to precisely manipulate local strain in 2D materials by gaining a profound understanding of their mechanical behavior when interfaced with various substrates. The first configuration that we consider involves an intricate examination of the geometric variations of 2D materials, incorporating a two-dimensional array of hard-wall rounded nanopillars (as depicted in Fig.~\ref{fig:Fig1}(a)). The intricate surface profile of each pillar, characterized by a half-semicircular structure with height equal to the radius, that is, $h=r$ (as depicted in Fig.\ref{fig:Fig1}(c)), is expertly captured. The second configuration features a rectangular one-dimensional substrate grating overlaid with a 2D material filling in the formed trenches, as visualized in Fig.~\ref{fig:Fig1}(b). Remarkably, the inherent complexity of these configurations is elegantly distilled into a concise 1D formulation, as illustrated in both Fig.~\ref{fig:Fig1}(c) and Fig.~\ref{fig:Fig1}(d). Through this research, we provide a deeper understanding of the mechanical response of 2D materials, opening up new avenues for precision engineering applications.

\section{Theoretical Framework}
In this investigation, we present a rigorous theoretical framework to quantify the total energy per unit cell area of the patterned structure. This energy  is expressed as follows:
\begin{equation}
E = E_{\text{strain}} + E_{\text{bending}} + E_{\text{adhesion}}.
\end{equation}
where each component $E_{\text{strain}}$, $E_{\text{bending}}$, and $E_{\text{adhesion}}$, represents a distinct contribution originating from essential mechanical phenomena: strain, bending, and van der Waals adhesion energies, respectively. Detailed expressions for these distinct components can be found in the Appendices \ref{nanopillars} and \ref{trench}, corresponding to two-dimensional nanopillar array and one-dimensional grating configurations, respectively.

To assess the strain experienced by the 2D material, we utilize the following equation:
\begin{equation}
    \varepsilon(x) = \sqrt{1+y'(x)^2}-1.
    \label{strain1}
\end{equation}
We assumed here that the area of 2D material equals the unit cell area of $2a\times 2a$ and no sliding is allowed, such that any tilt of 2D material results in the local strain given by Eq.~(\ref{strain1}).

\section{Results and discussion}
\begin{figure*}[t! ]
\includegraphics[height=8.5cm]{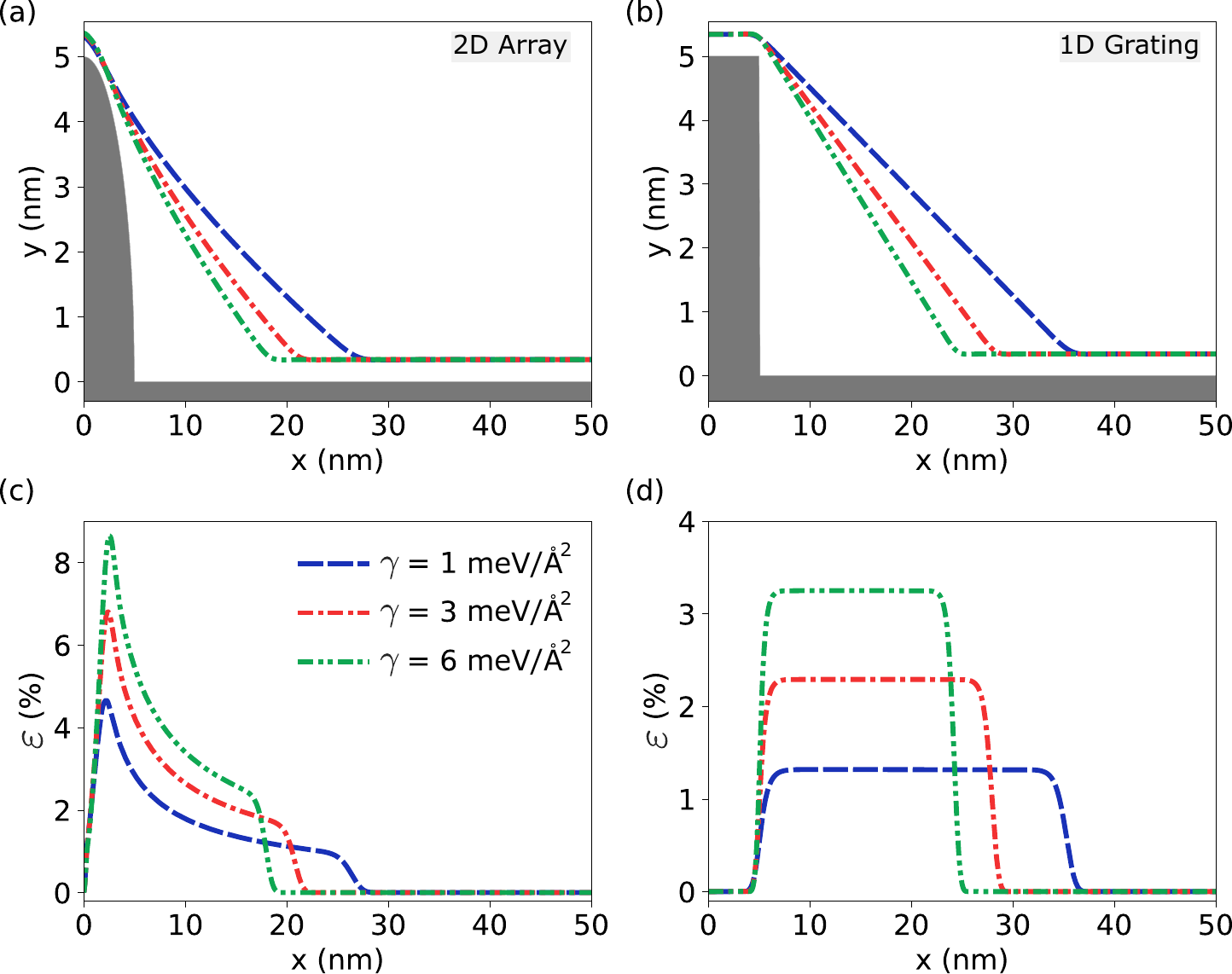}
\caption{\label{fig:Fig2} Engineering MoSe$_2$ structure on two-dimensional nanopillar array and one-dimensional grating configurations. (a) Demonstrates the customized behavior of MoSe$_2$ on nanopillar array, while (b) illustrates the controlled modulation of the MoSe$_2$ structure on grated substrate with varying interfacial adhesion energy ($\gamma$). (c) Unveils the intricate strain distribution within the MoSe$_2$ structures on nanopillars, and (d) presents the comprehensive strain distribution in the MoSe$_2$ structures in the grated substrates over a range of $\gamma$ levels. These figures collectively highlight the profound impact of the interfacial adhesion energy ($\gamma$) on MoSe$_2$ positioning and the distribution of strain. A constant value $a$ of 50 nm is maintained in all configurations to ensure a consistent basis for comparison. The gray area in (a) and (b) represents the nanostructured substrate.}
\end{figure*}

Figure~\ref{fig:Fig2} illustrates the mechanical response of MoSe$_2$, which serves as a representative example of 2D materials when strategically placed on a two-dimensional array of nanopillars or on a rectangular one-dimensional substrate grating. Since the adhesion energy can be engineered by choosing a 2D material and substrate, we present our results for an experimentally accessible range of adhesion energies $\gamma$.

In Figure~\ref{fig:Fig2}(a), we observe the precise tailoring of MoSe$_2$ behavior when placed on a nanopillar array. This figure provides visual evidence of van der Waals interactions that govern the material's positioning atop the nanopillars. Manipulation of the MoSe$_2$ geometry on one-dimensional substrate grating with varying values $\gamma$ is demonstrated in Figure~\ref{fig:Fig2}(b). This highlights the versatility of MoSe$_2$ in adapting to different interfacial adhesion energies.

Figure~\ref{fig:Fig2}(c) provides an in-depth exploration of the strain distribution within the MoSe$_2$ structures on nanopillars. It becomes evident that the strained area of the 2D material can be manipulated by varying the interfacial adhesion energy. 
As $\gamma$ increases or decreases, the extent of the strain within the material adjusts accordingly, leading to a controlled modulation of the strained area. This tunability is of paramount importance in tailoring the mechanical response of 2D materials for specific applications, particularly in SPEs applications.

In Figure~\ref{fig:Fig2}(d), we extend a similar analysis to MoSe$_2$ structures within grating configurations, providing a detailed view of its response to varying $\gamma$ values. Interestingly, a distinct behavior is observed in grated structures: the strain in the strained area remains relatively constant until the 2D material fully contacts the substrate in the trenches. This phenomenon suggests that grating configurations exhibit a unique mechanical response characterized by a sudden change in strain behavior upon contact, distinct from the gradual modulation observed in nanopillar arrays.

\begin{figure*}[htb]
\includegraphics[height=8.5cm]{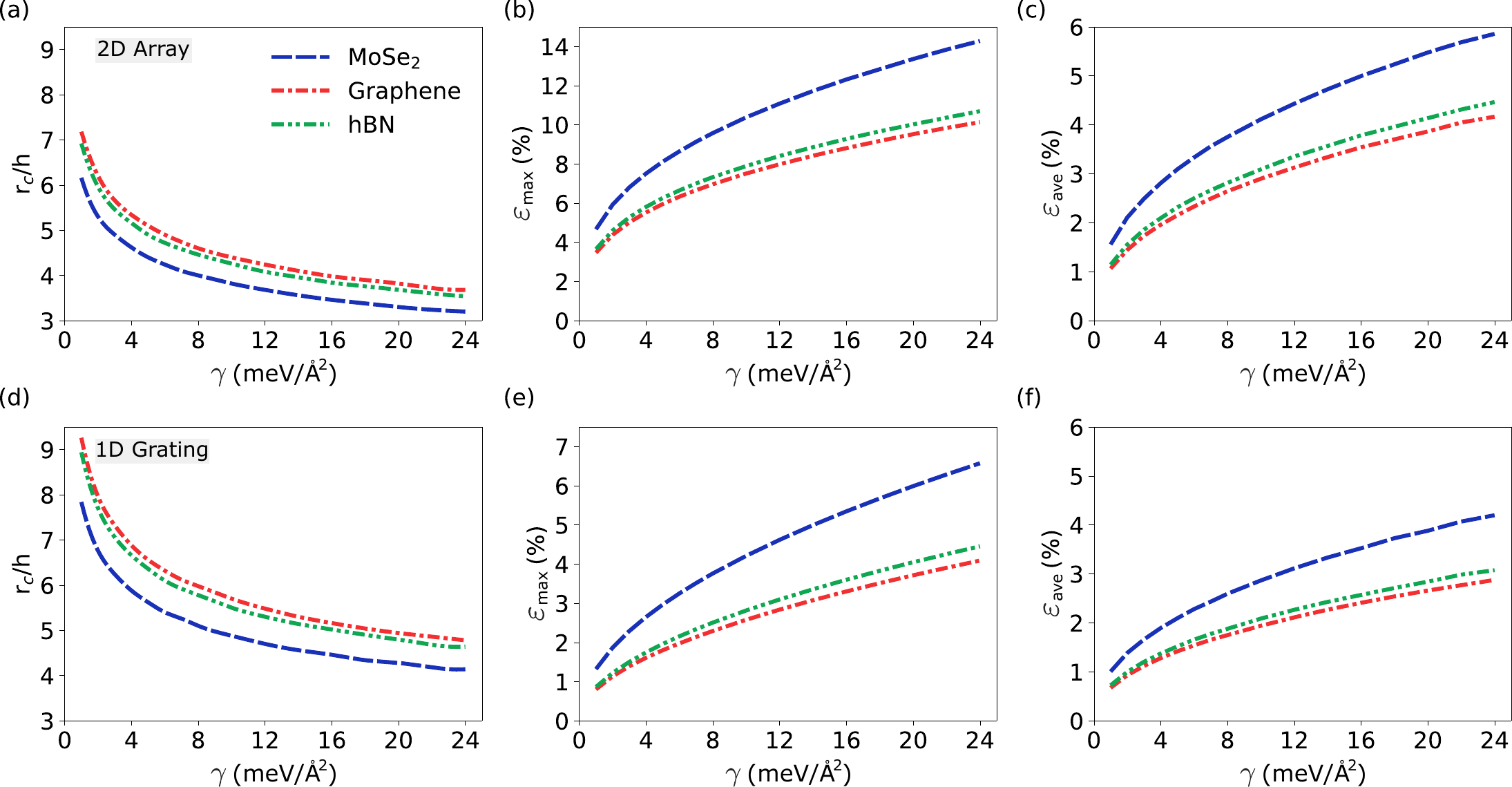}
\caption{\label{fig:Fig3} The evolution of the critical distance ratio ($r_c$) relative to the heights of the nanopillars and the depths of the grating in response to the variations $\gamma$ between different 2D materials and substrates is presented in (a) and (d), respectively. The maximum strain within the 2D material is analyzed in (b) for nanopillar array configurations and (e) for grating configurations as a function of the interfacial adhesion energy. The average strain within the 2D material is illustrated in (c) for the nanopillar array and in (f) for grating scenarios.
}
\end{figure*}

To dig deeper into the response of 2D materials placed in nanopillar arrays and patterned substrates, we introduce the concept of a critical distance, denoted $r_c$. This critical distance represents the point at which the strain levels approach near-zero values. Figure~\ref{fig:Fig3} presents the ratio of the critical distance $r_c$ to $h$, together with the maximum and average strain values as functions of $\gamma$.

Figs.~\ref{fig:Fig3}(a) and~\ref{fig:Fig3}(d) depict the ratio of $r_c/h$ versus interfacial adhesion energy for the case of the nanopillar array and the patterned substrate, respectively. In particular, as $\gamma$ increases, this ratio decreases consistently. Importantly, this trend remains consistent over varying periods. Furthermore, the ratio exhibits variations between different materials, primarily due to differences in their Young modulus and bending stiffness, as detailed in Table~\ref{tab:table1}.

The evolution of maximum strain as a function of $\gamma$ is depicted in Figs.~\ref{fig:Fig3}(b) and~\ref{fig:Fig3}(e) for nanopillar array and the patterned substrate, respectively. It becomes evident that as $\gamma$ increases, the maximum strain exhibits a corresponding increase, due to the changes in the shape of the 2D material induced by stronger adhesion between its layers. Additionally, it should be noted that the maximum strain for 2D materials on the nanopillar arrays is approximately twice that observed for 2D materials atop one-dimensional patterned substrates.


Since exciton emission can occur from different locations in 2D material, we introduce the concept of average strain as follows:
\begin{equation}
    \varepsilon_{ave}=\dfrac{1}{r_c}\int_0^{r_c}\varepsilon(x)dx,
\end{equation}
and illustrate it as a function of $\gamma$ for different materials in Figs.~\ref{fig:Fig3}(c) and~\ref{fig:Fig3}(f) for nanopillar array and rectangular grating of the substrate, respectively. Evidently, the average strain exhibits an upward trajectory with increasing interfacial adhesion energy. However, it should be noted that, in contrast to the differences in maximum strain, the average strain values for nanopillar arrays and grating configurations are observed to be in close proximity to each other.

\begin{figure*}[htb]
\includegraphics[height=8.5cm]{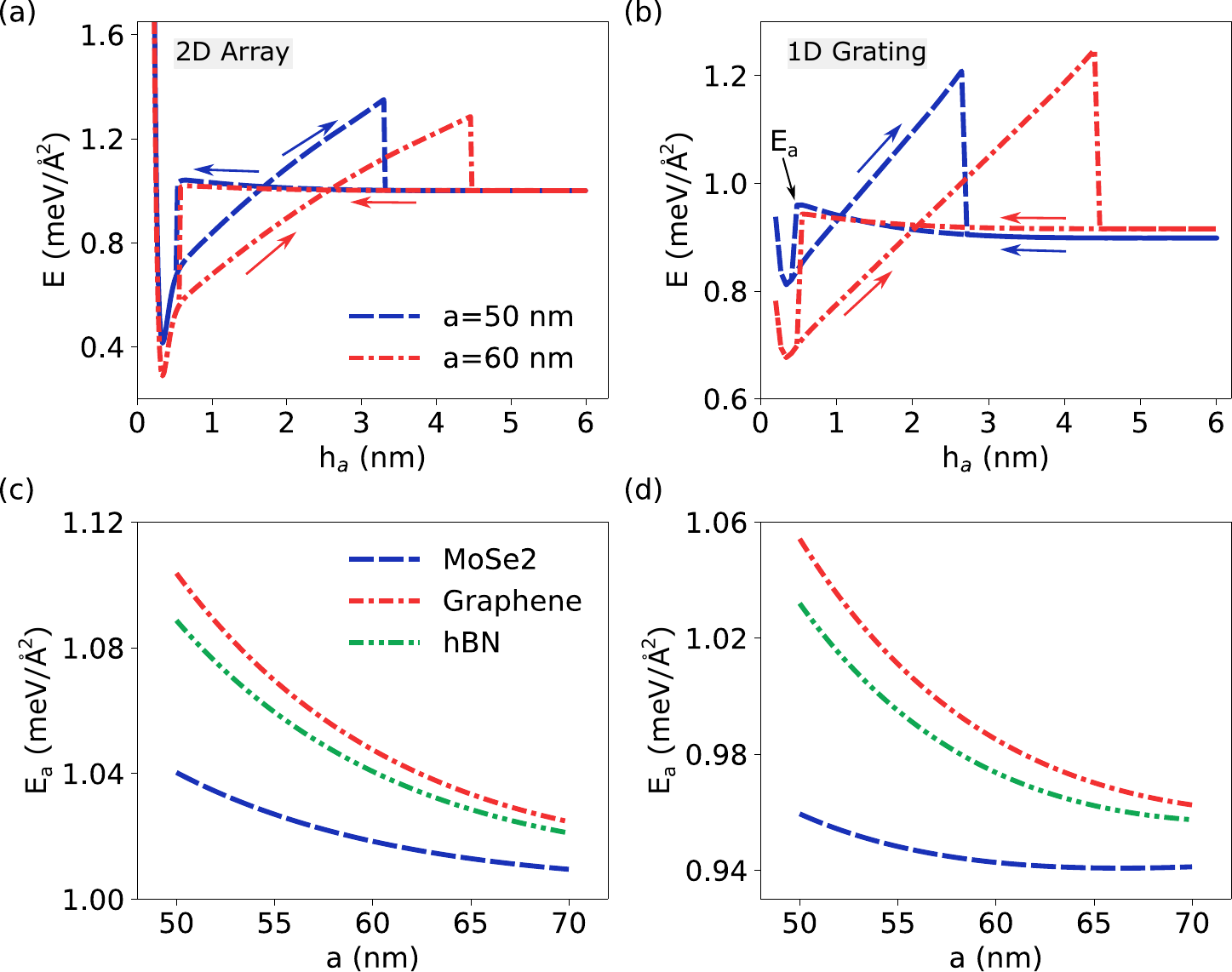}
\caption{\label{fig:Fig4} Energy hysteresis and analysis of the activation energy. (a) and (b) show the energy hysteresis for the nanopillar array configuration and the grated substrate, respectively, as a function of $h_a$ for different lengths of the period. (c) and (d) show the activated energy for the nanopillar array configuration and the grated substrate configuration, respectively, as a function of periods across diverse materials. The adhesion energy is set to 1 meV/$\si{\angstrom}^2$ in both cases. These comprehensive analyses provide deep insight into the energy behavior of these structures across various parameters.
}
\end{figure*}
Figure \ref{fig:Fig4} presents a detailed analysis of the energy variations within the sample structures. Specifically, Figs. \ref{fig:Fig4} (a) and \ref{fig:Fig4} (b) provide information on the energy hysteresis observed with respect to changes in $h_a$. In particular, as the parameter $h_a$ increases, an intriguing phenomenon unfolds: The 2D material gradually disengages from the substrate, leading to a gradual increase in overall energy. However, a critical threshold of $h_a$ is reached, beyond which a remarkable event occurs, the complete detachment of the 2D material from the substrate. This detachment event is characterized by a sharp and pronounced drop in total energy. Beyond this critical threshold, further increases in $h_a$ no longer significantly impact the energy of the system as it stabilizes in a detached state. Conversely, when reducing $h_a$, the energy landscape exhibits a distinct behavior. Initially, as $h_a$ decreases, the energy remains relatively constant until it reaches a lower value. Subsequently, a departure from this equilibrium occurs, resulting in an increase in energy. Ultimately, the energy reaches the activation energy, denoted $E_a$, signifying the point at which the 2D material re-establishes contact with the substrate. This reengagement prompts a swift and substantial drop in energy as the structure strives to achieve its optimized configuration.

Turning our attention to Figs. \ref{fig:Fig4}(c) and \ref{fig:Fig4}(d), these plots provide a detailed analysis of the activation energy as a function of the period in a range of materials. A clear and consistent trend emerges: as the period increases, the activation energy exhibits a consistent decrease. This observation underscores the intriguing relationship between the period and the energy required for structural relaxation. Specifically, longer periods are associated with a reduced demand for energy during the relaxation process, a phenomenon of significance in the study of material behavior.

\section{Conclusion}
In summary, our advanced theoretical framework provides a comprehensive understanding of the mechanics of 2D material membranes when transferred to different patterned substrates, such as nanopillar array and grating structures. It elucidates the critical role of the interfacial adhesion energy ($\gamma$) in the governing behavior of these materials. As $\gamma$ increases, significant alterations are observed in the strain, critical period, and energy profiles, highlighting the sensitivity of 2D materials to this parameter and offering the prospect of precise engineering for customized nanoscale applications.
A notable outcome of this study is the concept of a critical aspect ratio of the height of the nanostructure to the period ($r_c$), which emerges as a key metric that characterizes the response of the system. $r_c$ varies with both $\gamma$ and the specific material, providing valuable information on how 2D materials adapt and interact with different substrate environments.

The analysis of energy hysteresis reveals intriguing detachment and reengagement events as the height of the 2D material on the substrate changes. These events are reflected in sharp fluctuations in energy profiles, emphasizing the dynamic nature of these systems and the energy barriers associated with structural rearrangements.
Furthermore, the investigation of activation energy demonstrates that it decreases with a larger period, shedding light on the role of the pattern period in shaping material behavior. This advanced framework not only enhances our understanding of 2D material mechanics but also presents opportunities for precision engineering, paving the way for tailored nanoscale materials and devices with specific mechanical properties, thus contributing to the evolving field of 2D materials and their potential technological applications.

\appendix
\section{Energy Contributions}
The geometric profile of the 2D material overlaid on nanopillar arrays and grating structures can be described through a Fourier series, as defined by:
\begin{equation}
y(x) = A_0 + \sum_{n=1}^{N} A_{n} \cos\left(\frac{n\pi x}{a}\right).
\end{equation}
Upon enforcing boundary conditions, this representation simplifies to:
\begin{equation}
y(x) = h_a + \sum_{n=1}^{N} A_{n} [\cos(\dfrac{n\pi x}{a}) - \cos (n\pi)].
\end{equation}

Here, $h_a$ denotes the height of the 2D material at $x=a$ (see Fig.~\ref{fig:Fig1}), while $2a$ represents the period of the structure, and $A_n$ signifies the Fourier coefficient. In this section, we discuss the components that contribute to the total energy of the 2D profile, encompassing the stresses, bending, and van der Waals adhesion energies.

\subsection{2D Materials on Nanopillar Arrays} \label{nanopillars}
To elucidate the underlying energy expressions that govern this configuration, we present the following equations.

The quantification of strain energy is expressed as:
\begin{equation}
E_{\text{strain}} = \pi Y \int_{0}^{a} x \varepsilon(x)^2 dx,
\end{equation}
where $Y$ represents the Young's modulus.

\begin{table}
\caption{\label{tab:table1} Mechanical Characteristics of Various Materials, including Young's Modulus ($Y$), Bending Stiffness ($\chi$), and Monolayer Thickness ($d$)~\cite{yang2017brittle, lee2008measurement}.}
\begin{ruledtabular}
\begin{tabular}{c c c c}
        \textbf{Materials} & \textbf{Y (GPa)} & \textbf{$\chi$ (eV)} & \textbf{$d$ (nm)} \\
\hline
         MoSe$_2$ & 177 &  9.96 & 0.700\\  
        Graphene & 1000 & 1.40 & 0.335   \\
        hBN & 865 & 0.86 & 0.334  \\
\end{tabular}
\end{ruledtabular}
\end{table}

The energy associated with bending is articulated as:
\begin{equation}
E_{\text{bending}} = \pi \chi \int_{0}^{a} x \frac{y''(x)^2}{(1+y'(x)^2)^{3}} dx,
\end{equation}
with $\chi$ denoting the bending stiffness. The values for $Y$ and $\chi$ in various materials are provided in Table~\ref{tab:table1}. 

To evaluate adhesion energy, we employ the Morse potential~\cite{morse1929} between layers:
\begin{equation}
\begin{aligned}
E_{\text{adhesion}} = & \pi 4a^2 \gamma + (4-\pi)a^2 U_{2D}(y(a))\\
& +2\pi\int_0^a x~U_{2D}(y(x))dx,
\end{aligned}
\end{equation}
where $\gamma$ represents the interfacial adhesion energy between the 2D material and the substrate. The Morse potential, $U_{2D}(y)$, is given by:
\begin{equation}
U_{2D}(y) = \gamma \left(e^{-2\sqrt{5}((y/h_0)^2-1)}-2e^{-\sqrt{5}((y/h_0)^2-1)}\right).
\end{equation}
Here, $h_0=0.34~nm$ stands for the equilibrium distance between layers. Within these equations, $y'(x)$ and $y''(x)$ represent the first and second derivatives of the shape of the 2D material, respectively. This comprehensive ensemble of energy components collectively serves to elucidate the intricate mechanical interactions between 2D materials and nanopillar arrays.

\subsection{2D Material on rectangular grating substrate} \label{trench}
Another intriguing configuration involves positioning 2D materials over a grating structure. The fundamental energy expressions governing this configuration are described in detail below. 

The strain energy can be obtained by:
\begin{equation}
E_{\text{strain}} =  Ya \int_{0}^{a} \varepsilon(x)^2 dx.
\end{equation}

To calculate the bending energy, we employ:
\begin{equation}
E_{\text{bending}} = a\chi \int_{0}^{a} \frac{h''(x)^2}{(1+h'(x)^2)^{3}} dx.
\end{equation}

Finally, we calculate the adhesion energy using:
\begin{equation}
    \begin{aligned}
        E_{\text{adhesion}} = & 4a^2\gamma + 4a\int_{0}^{a}U_{2D}(y(x))dx\\
        &+4a\int_{0}^{a}dx\int_{-r}^{r}\bigl(U_{1D}(x-x',y(x)-r)\\
&-U_{1D}(x-x',y(x))\bigr)dx'\\
&+4a\int_{0}^{a}dx\int_{0}^{r}U_{1D}(x-r,y(x)-y')dy',
\end{aligned}            
\end{equation}
where $U_{1D}(x,y)$ is described as:
\begin{equation}
    \begin{aligned}
        U_{1D}(x,y) = & \gamma \frac{2\times 5^{1/4}}{\sqrt{\pi}h_0}\\
        &\left(\frac{1}{\sqrt{2}}e^{-2\sqrt{5}((x^2+y^2)/h_0^2-1)}-e^{-\sqrt{5}     ((x^2+y^2)/h_0^2-1)}\right).
    \end{aligned}
\end{equation}
\begin{acknowledgments}

We gratefully acknowledge the support of the Air Force Ofﬁce of Scientiﬁc Research under award number FA9550-22-1-0312 and the computational facilities of the Center for Computational Research at the University of Buffalo (\url{http://hdl.handle.net/10477/79221}).  J.R.H. acknowledges support from the Air Force Office of Scientific Research (Program Manager, Dr. Gernot Pomrenke) under award number FA9550-20RYCOR059.

\end{acknowledgments}

\bibliographystyle{apsrev4-1}
\bibliography{References1}

\end{document}